# Development of an ELT XAO testbed using a Mach-Zehnder wavefront sensor: calibration of the deformable mirror

Christian Delacroix*, Maud Langlois, Magali Loupias, Eric Thiébaut, Louisa Adjali, Jonathan Leger and Michel Tallon

CRAL, Observatoire de Lyon, CNRS UMR 5574, Université Lyon 1, 9 avenue Charles Andrée, 69230 Saint-Genis Laval, France

## ABSTRACT

Extreme adaptive optics (XAO) encounters severe difficulties to cope with the high speed (>1kHz), high accuracy and high order requirements for future extremely large telescopes. An innovative high order adaptive optics system using a self-referenced Mach-Zehnder wavefront sensor (MZWFS) allows counteracting these limitations. This sensor estimates very accurately the wavefront phase at small spatial scale by measuring intensity differences between two outputs, with a $\lambda/4$ path length difference between its two legs, but is limited in dynamic range due to phase ambiguity. During the past few years, such an XAO system has been studied by our team in the framework of 8-meter class telescopes. In this work, we report on our latest results with the XAO testbed recently installed in our lab, and dedicated to high contrast imaging with 30m-class telescopes (such as the E-ELT or the TMT). After reminding the principle of a MZWFS and describing the optical layout of our experiment, we will show the results of the assessment of the woofer-tweeter phase correctors, i.e., a Boston Micromachine continuous membrane deformable mirror (DM) and a Boulder Nonlinear Systems liquid crystal spatial light modulator (SLM). In particular, we will detail the calibration of the DM using Zygo interferometer metrology. Our method consists in the precise measurement of the membrane deformation while applying a constant deformation to 9 out of 140 actuators at the same time. By varying the poke voltage across the DM operating range, we propose a simple but efficient way of modeling the DM influence function using a Gaussian model. Finally, we show the DM flattening on the MZWFS allowing to compensate for low order aberrations. This work is carried out in synergy with the validation of fast iterative wavefront reconstruction algorithms, and the optimal treatment of phase ambiguities in order to mitigate the dynamical range limitation of such an MZWFS.

## 1. INTRODUCTION

The direct detection and characterization of exoplanetary systems and debris disks around nearby stars requires achieving high-contrast imaging capabilities, very close to the host star. Recently deployed new instruments (Palomar/P1640,[1] Gemini/GPI,[2] and VLT/SPHERE[3]) use extreme adaptive optics (XAO) to compensate for the atmospheric turbulence, and coronagraphy to attenuate the signal of the host star. In particular, low-order aberrations such as tip-tilt due to jitter or imperfect star centering highly degrade coronagraphic performance, and can be evaluated by various existing sensing techniques[4]. Following recent on-sky tests, the centering of the star image on the coronagraph is achieved with pointing errors smaller than 0.5 mas rms on SPHERE.[5] Commonly used wavefront sensors (WFS) such as Shack-Hartmann and Curvature are very robust and flexible, but poorly suited to high sensitivity wavefront measurements. For these conventional WFS concepts, the sensitivity reaches its minimum at a given distance from the optical axis and increases closer to the PSF core[6]. They suffer from the noise propagation effect, i.e., a poor sensitivity to low-order Zernike terms. Other WFS concepts exist, that maintain a very high sensitivity, constant at all separations, across a wide range of spatial frequencies, such as Mach-Zehnder WFS[7–9], Zernike WFS[10,11], or pyramid WFS[12]. For these WFS concepts, the noise propagation is low, and both low-order and high-order terms can be corrected very efficiently. High sensitivity wavefront measurements are mandatory to detect Earth-like planets with future extremely large telescopes (ELTs). For these ELTs, small inner working angle coronagraphs are envisioned (e.g. E-ELT/METIS[13]) such as vortex phase masks, which already equip 10m class ground-based telescopes (VLT/NACO[14,15], VLT/VISIR[16,17], LBT/LMIRCam[18], Subaru/SCExAO[19], and Keck/NIRC2[20]).

*christian.delacroix@univ-lyon1.fr; phone +33 478868547; cral.univ-lyon1.fr

In this paper, we present the setup and calibration of an experimental test bench at CRAL (Centre de Recherche Astrophysique de Lyon), dedicated to the study of XAO for 30m-class ELTs, which will require more than 20000 actuators[21]. Our test bench implements a Mach-Zehnder Wavefront Sensor (MZWFS) as shown in Fig. 1. In Section 2, we describe the characterization of the phase correctors, using a Zygo interferometer. In Section 3, we focus on the calibration of the deformable mirror (DM) for which we use a simple but effective method of modeling the influence function for all the DM actuators using a Gaussian function. The theoretical DM correction for a few Zernike modes is also presented. Finally, in Section 4, we show our first results of DM flattening and low order aberrations compensation with the MZWFS.

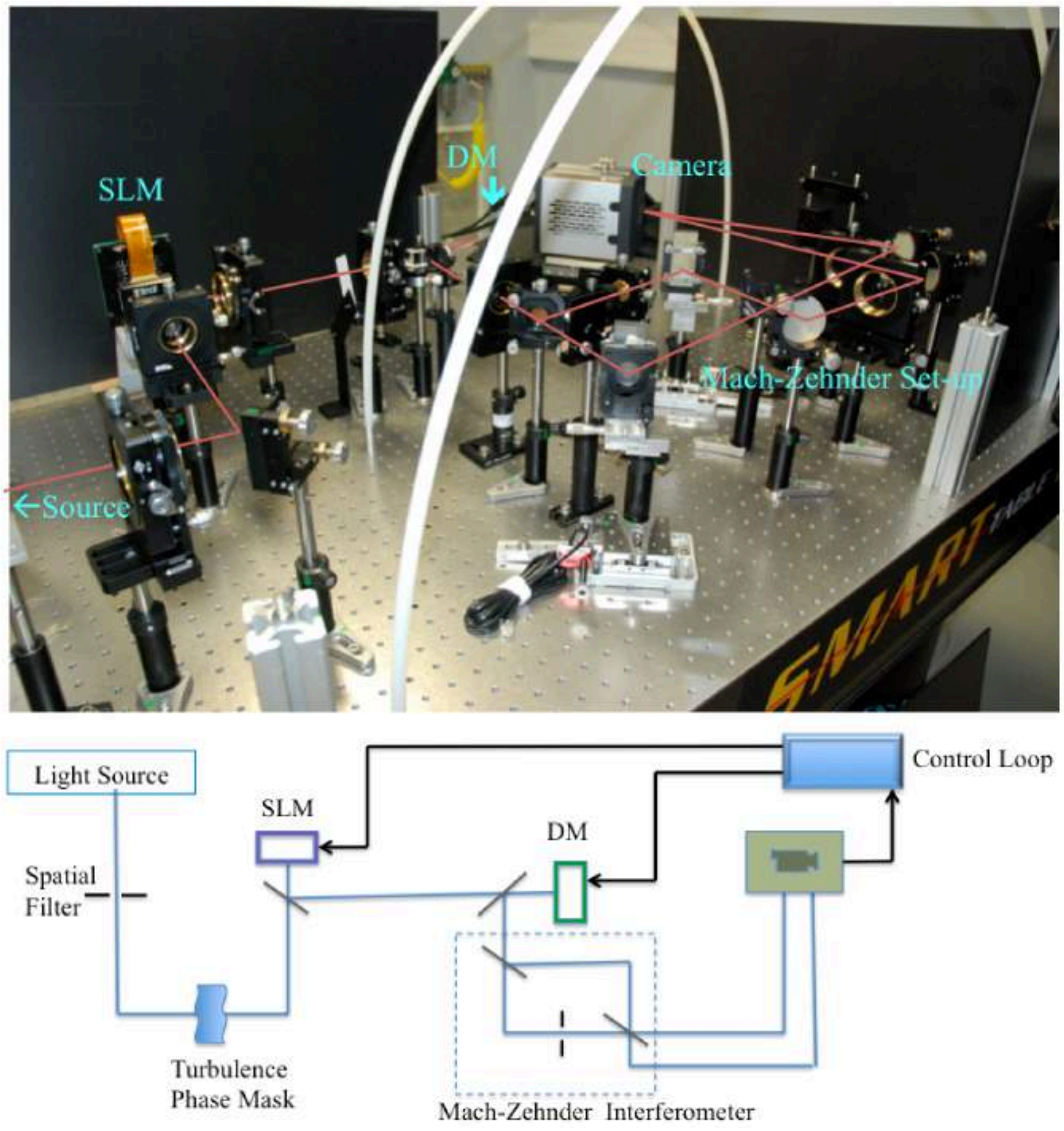

**Figure 1.** Optical setup of the experiment, including turbulence generator (to deform the wavefront), phase correctors (deformable mirror and spatial light modulator), and a MZWFS which forms a square of 0.2 m by 0.2 m. A beam splitter produces two copies of the same wavefront. One of the copies is spatially filtered and interferometrically combined with the unfiltered wavefront. Wavefront phase is transformed into intensity variations in the two pupil images produced by the interferometer.

## 2. ZYGO INTERFEROMETER METROLOGY OF THE PHASE CORRECTORS

A woofer-tweeter architecture is used in order to deliver the required high Strehl ratio (>95%). The woofer is a Boston Micromachine continuous membrane deformable mirror (DM), 140 actuators, 3.5 µm maximum stroke, 4.4×4.4 mm clear aperture (see Fig. 2). The tweeter is a Boulder Nonlinear Systems (BNS) liquid crystal spatial light modulator (SLM), 512×512 pixels on a 7.68 mm square grid, inducing phase modulation on a polarized input beam (see Fig. 3).

The flattening of the phase correctors was carried out using an industry-standard ZYGO's Verifire™ laser Fizeau interferometer, together with the following components:

(i) CCD camera 640 x 480 pixels, 104mm, zoom ×1 to ×5, measurement precision λ/20;

(ii) Thorlabs SI500 Shear plate assembly for Ø=25.4×Ø50mm beams;

(iii) Achromatic lenses L1 f=120mm D=50mm and L2 f=500mm D=60mm;

(iv) Newport optical table with auto-leveling pneumatic isolators.

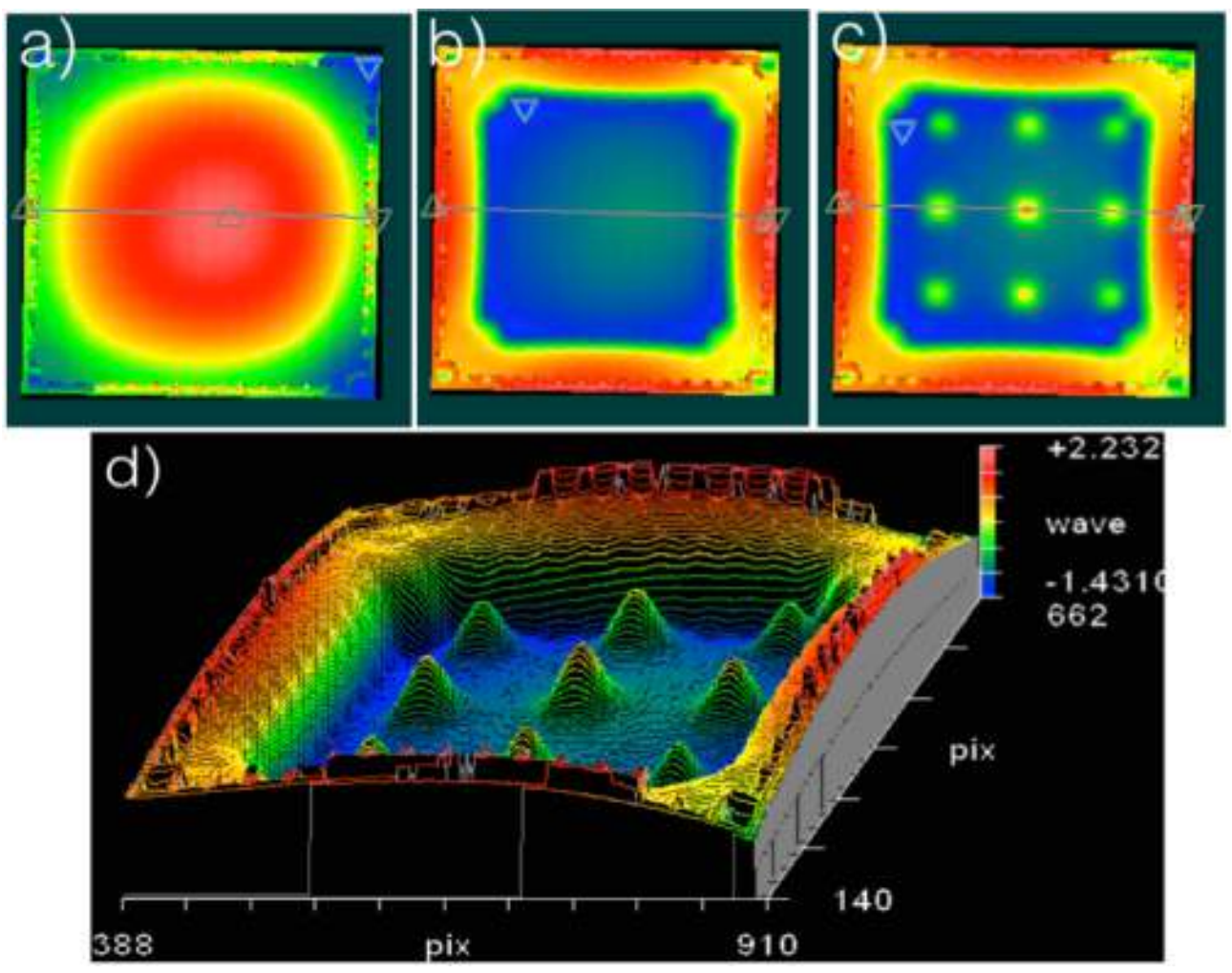

**Figure 2.** Profile of the DM membrane: a) mirror at rest ~1250nm deflection; b) manufacturer flatmap ~300nm deflection at 150V; c) & d) –135V offset applied to 9 actuators (1 map out of 16).

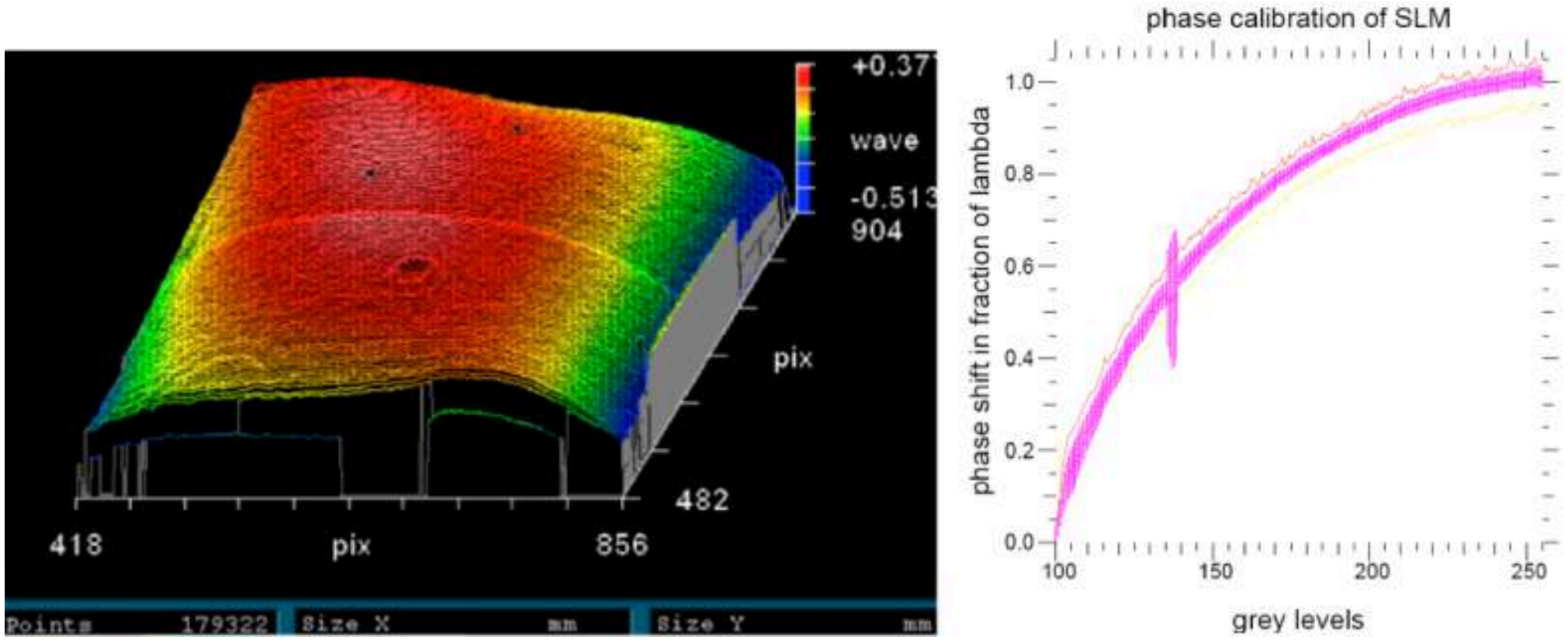

**Figure 3.** Left: Profile of the flattened SLM, using manufacturer flatmap. The central region has a residual deflection of ~50nm. Right: Response of the SLM, with a 2π-phase shift over 150 levels (a few nm phase resolution). The artifact on the curve is due to a dust particle trapped inside the Zygo.

## 3. CALIBRATION OF THE CONTINUOUS MEMBRANE DM

### 3.1 Flatmap subtraction

Dynamic effects on optical wavefront caused by atmospheric turbulence can be corrected by a DM, controlled by the application of voltages on actuators that hold the mirror surface. Our continuous DM has 140 uniformly distributed actuators in a 12×12 square, the four corners being unusable. The geometry is shown in Fig. 4 (top left). The inter-actuator spacing is 400 µm. The calibration of the DM requires assessing the effect of individual actuators separately. However, in the case of a continuous DM, the application of a known tension (called poke voltage) on one of the actuators results in a certain force inducing a finite displacement in the actuator vicinity, which is called the inter-actuator coupling. In our experiment, we apply a poke voltage to 9 actuators at the same time, for 16 different positions of the selected actuators, in order to cover the whole 140 actuators (see Fig. 4, top right). The selected actuators are sufficiently spaced (by a 4-actuator distance) and are almost not affected by the inter-actuator coupling, as we show later in section 3.2 about the influence function.

The DM is a Micro-Electro-Mechanical-System (MEMS), which offers a strong potential for miniature and cost, ideal for test benches. As an electrostatic mirror, it requires applying a voltage to deform the membrane. When the tension is set to 0 V at the DM rest position, one can only pull on the actuators. In order to form any kind of wavefront requiring push and pull on the actuators, we define a new zero position of the flat DM (called the flatmap) by setting all actuators to their mid-stroke at approximately 150 V (maximum tension is 215 V). We then apply different offsets within the specification limits of the DM, ranging from -135 V to +53 V (see Fig. 4, bottom).

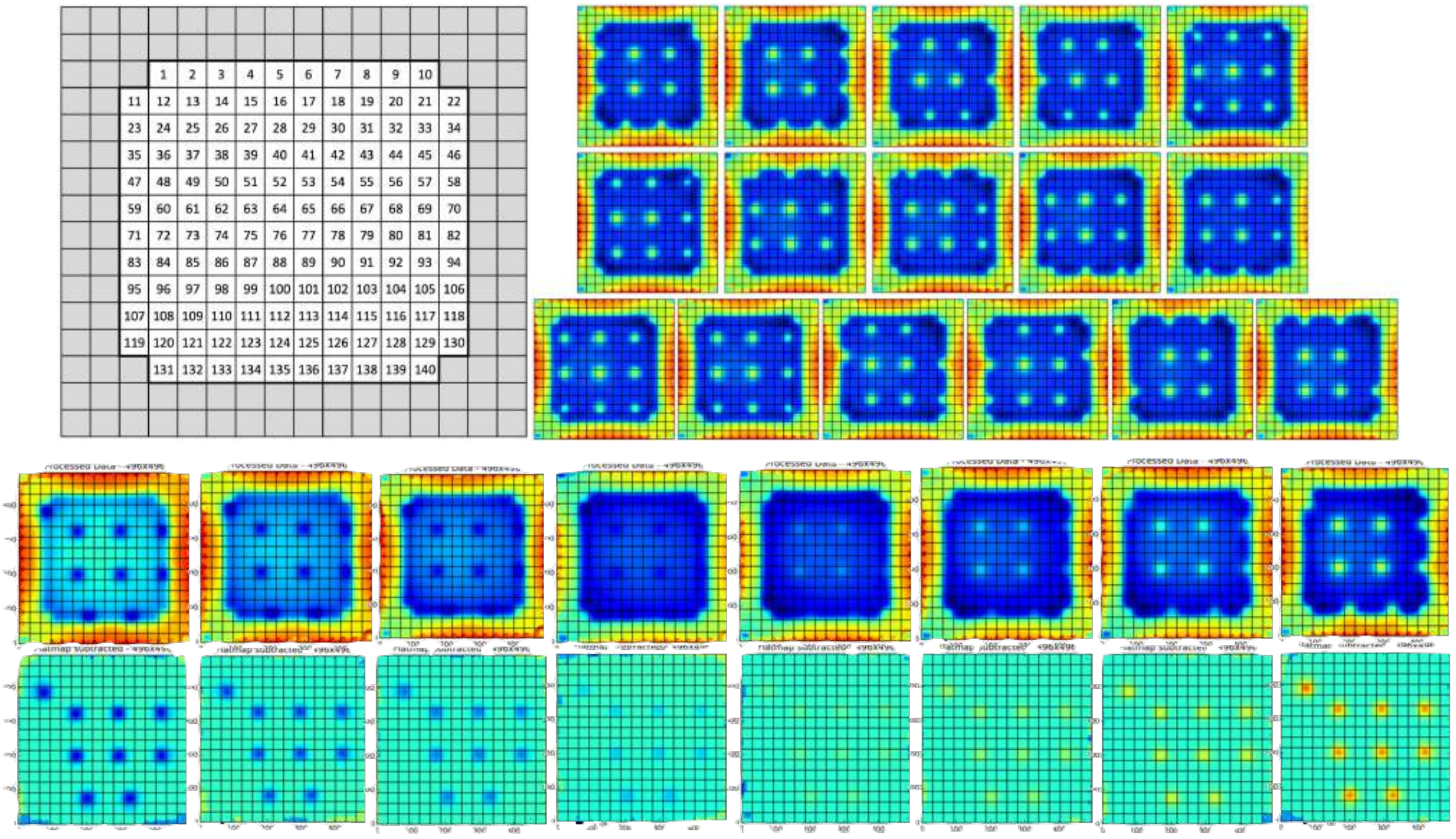

**Figure 4.** Top left: Actuator Map, with 12×12 actuators, the four corners being unusable. Grey region shows the non-deformable part of the mirror. Top right: The measurements have been made on 16 different positions (or maps) of 9 active actuators. Bottom: Set of figures showing the DM deformation for one specific map of 9 active actuators, with 8 different tension offsets. Negative offsets result from applying the 150 V flatmap. Top row: DM raw deformation as measured with the ZYGO interferometer. Bottom row: same results, after subtraction of the flatmap.

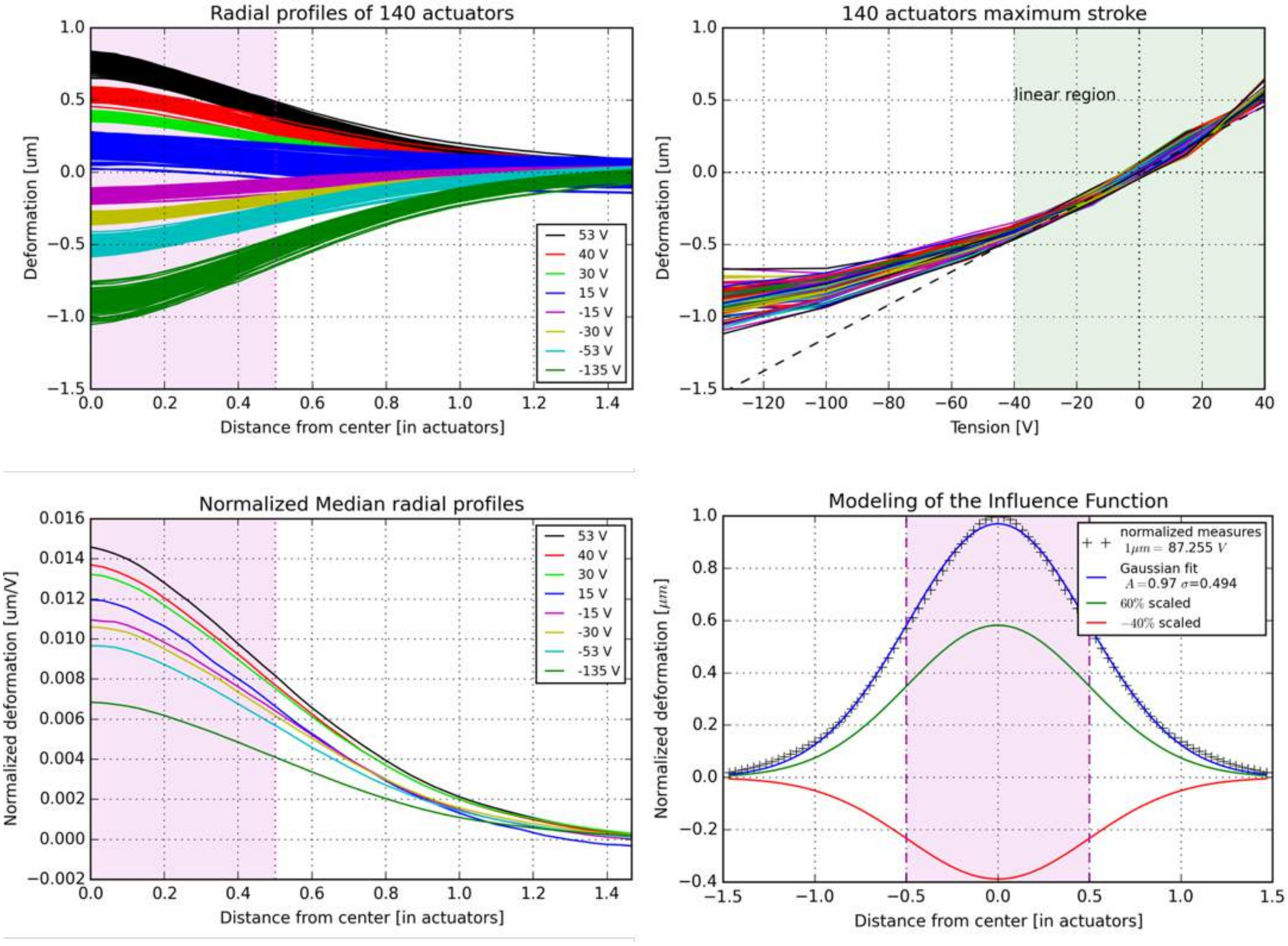

**Figure 5.** DM calibration. Top left: Absolute deformation radial profile for each actuator individually, at various tension offsets ranging from –135 V to +53 V. Top right: Deflection curve for each actuator peek deformation, function of tension offsets. Bottom left: Same as top left, medianed and normalized by the tension offsets. Bottom right: Modeling of the DM influence function with a 2D Gaussian model (blue curve). The green and red curves show examples of +60% and –40% scaled deformation, respectively.

### 3.2 Modeling of the influence function

Our experiment allows us to study the individual response of the 140 actuators, for a set of 8 different tension offsets ranging from –135 V to +53 V. For each of these, we observe the radial profile of the membrane deformation, after subtracting the associated flatmap, as shown in Fig. 5 (top left). We measure a 13% inter-actuator coupling, in perfect agreement with the manufacturer specifications. We notice that this inter-actuator coupling mainly affects the actuators directly adjacent to the active one, and is negligible (<0.05%) at a 4-actuator distance. In order to study the relationship between tension and deformation, we observe the peek deformation (maximum stroke) of all 140 actuators (Fig. 5, top right), function of the poke voltage on either side of the flatmap position (i.e. 0 V offset). The figure shows that the actuators respond quasi-linearly in the –40 V to +40 V offset region, allowing a deformation range larger than 1 µm, which is more than needed for our application.

We also calculate the normalized median of 140 actuators for each tension offset (see Fig. 5, bottom left). The resulting deformation profiles within the –40 V to +40 V offset region are quite similar, with a ~12 ± 1 nm/V peek deformation. From these results, we infer a best-fit model for the measured influence function of all the actuators. Different models are used amongst the literature.[22–24] Taking the median of the normalized radial profiles within the linear region (–40 to +40 V), we fit the DM influence function with a 2D Gaussian model (see Fig. 5, bottom right) possessing the following parameters: amplitude = 0.97 and standard deviation = 0.494.

### 3.3 Theoretical DM correction: Yorick code

This section shows how we compute the DM correction, to compensate for any wavefront aberration measured by our MZWFS. Using our Gaussian model described in the previous section, we calculate the influence function matrix that needs to be applied to the DM actuators in order to produce the measured phase errors. Phase errors can be approximated by a linear combination of Zernike polynomials. We show in Fig. 6 our simulated reconstruction of three Zernike polynomials (tip, defocus, and coma), providing a very accurate reproduction. Our simple algorithm (i.e. Zernike polynomial calculation and matrix inversion) is coded in Yorick language, which allows quick development, easy debugging and plotting functionalities. The choice of coding with Yorick is motivated by the use of this language in the fractal iterative method (FRiM), a fast iterative algorithm for minimum variance wavefront reconstruction and control on ELTs.[25] Yorick is also used in Yao, an adaptive optics simulation software package that has been used by the community for nearly 15 years.[26]

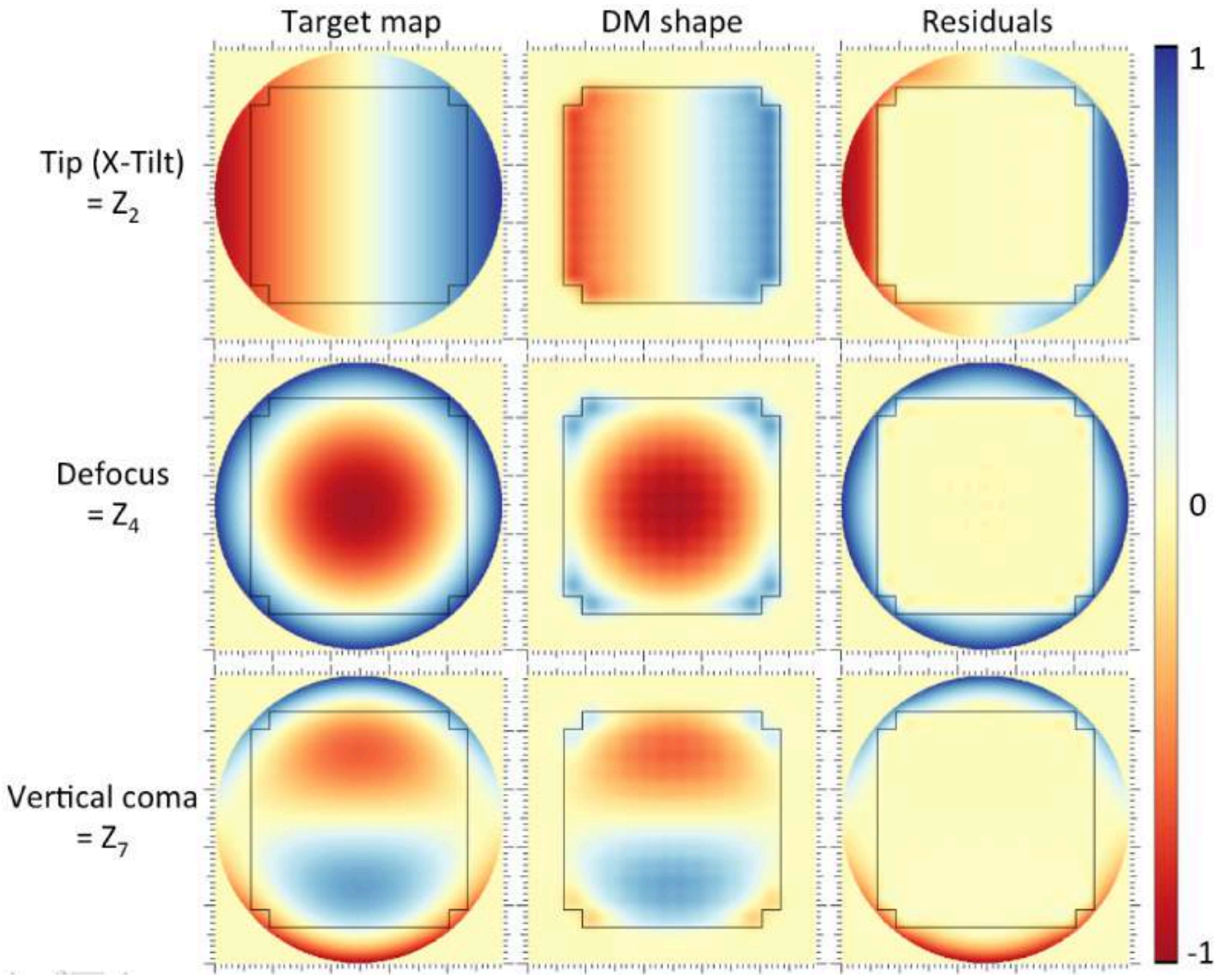

**Figure 6.** Reconstruction of three Zernike polynomials (tip, defocus, and coma) obtained by solving the inverse matrix problem, which associates the influence function matrix to the actuators poke voltage matrix (i.e. DM shape). Gaussian model described in the previous section, we calculate the influence function coefficients matrix that needs to be applied to the DM actuators in order to produce the measured phase errors.

## 4. DM FLATTENING AND LOW ORDER ABERRATIONS COMPENSATION

After having been measured and characterized using a Zygo interferometer, the DM was installed on the MZWFS testbed. The first experimental tests have been carried out. At the moment, the setup is equipped with a monochromatic light source. At the MZWFS output, two phase measurements ($I_1$ and $I_2$) are performed (see Fig. 7, top left). By combining the two outputs as $(I_2–I_1) / (I_2+I_1)$, we retrieve the initial phase aberrations.[27] This simple phase retrieval method is already very accurate for small phase errors, and allows to precisely calibrate the system. The DM is then controlled in closed loop to compensate for the phase aberrations (see Fig. 7, bottom left).

The pupil is sampled with ~1000 pixels per diameter (see Fig. 7, right). At the moment, the DM influence function is not yet implemented in the closed-loop automation, which explains why we have not yet reached the expected high sensitivity. Besides, a more sophisticated phase retrieval method might be needed to avoid the assumption of small phase errors. Although, the MZWFS communicates correctly with the DM allowing flattening and low order aberrations compensation.

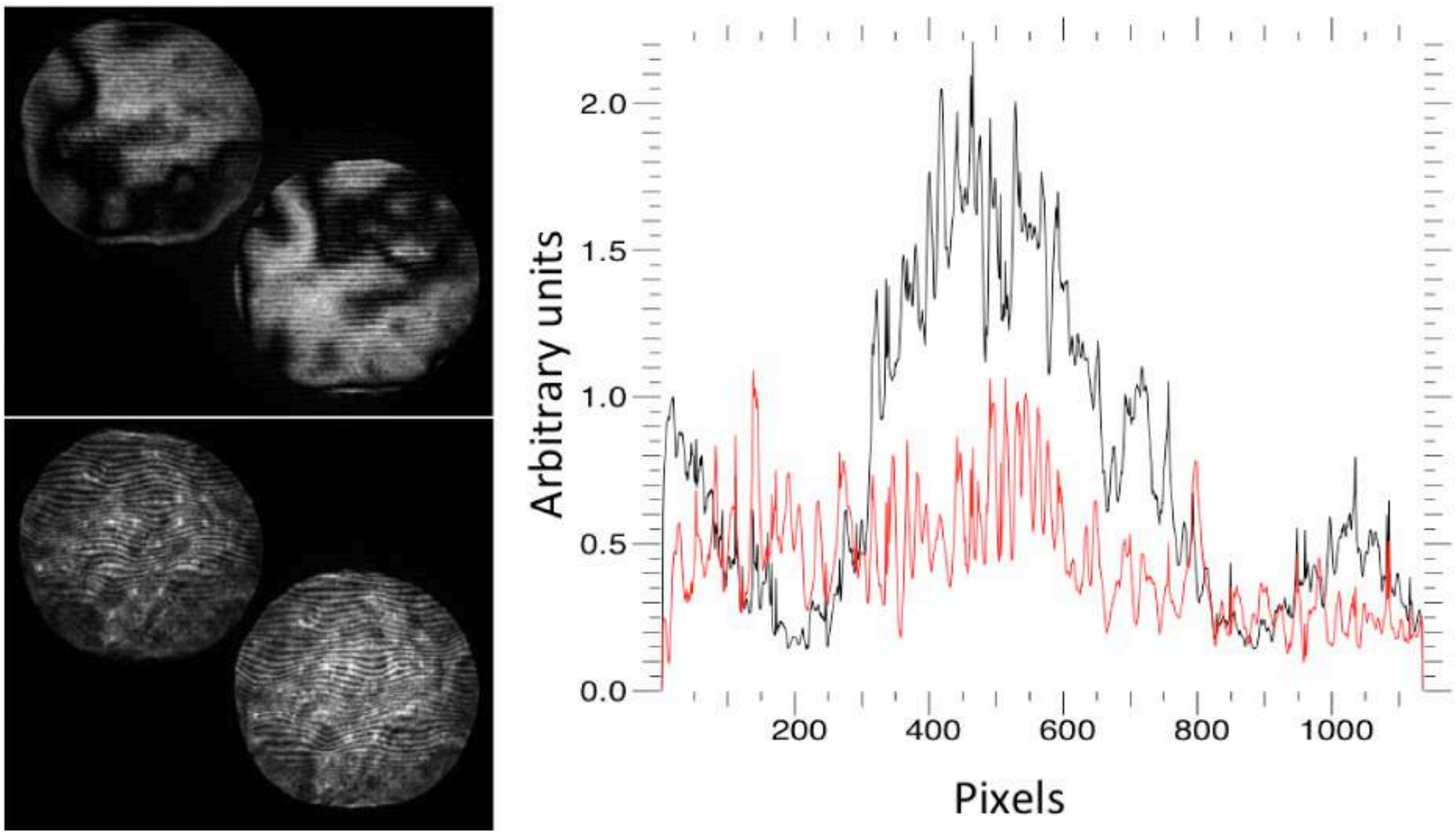

**Figure 7.** Left: MZWFS two output phase measurements, before (top) and after (bottom) a low order correction with the flattened DM, but without the implementation of the modeled influence function. Right: simple phase retrieval, where the *y* axis has no physical unit, as it only illustrates the comparison ratio between the two abovementioned cases (black curve = without DM correction, red curve = with DM correction). Note that when using monochromatic light these measurements are slightly impacted by a fixed narrow fringe pattern incoming from the Mach-Zehnder beam splitters themselves.

## 5. CONCLUSION AND PROSPECTS

We have built an experimental WFS dedicated to ELTs to accurately estimate and correct the wavefront phase errors at small spatial scale, using a Mach Zehnder interferometer and two phase correctors (DM and SLM) working in closed loop. This setup is an evolution of a previous XAO system studied by our team for 8m-class telescopes and cited in the introduction of the paper. The phase correctors were characterized using a Zygo interferometer, and the DM influence function was modeled using a 2D Gaussian function. A simple phase retrieval algorithm was coded using Yorick language. New results were obtained with the MZWFS using the characterized phase correctors, with monochromatic light. The results show however that improvements are still needed. More specifically, the Yorick influence function algorithm is to be integrated soon to the setup in order to precisely control the DM and compensate for the measured aberrations. A precise calibration of the SLM is also in progress.

Despite their weakness, our results demonstrate the proper alignment and operation of our MZWFS to be used in closed loop with phase correctors using monochromatic light. We are working to make the MZWFS ready to operate with polychromatic light, with the appropriate optical design. The next upgrade would be to equip the setup with a science path, allowing to test coronagraphic applications both in focal plane and pupil plane.

## ACKNOWLEDGEMENT

The authors are grateful to the LABEX Lyon Institute of Origins (ANR-10-LABX-0066) of the Université de Lyon for its financial support within the program "Investissements d'Avenir" (ANR-11-IDEX-0007) of the French government operated by the National Research Agency (ANR).